\documentclass[12pt]{article}
\usepackage{graphicx}
\begin{document}

version 28.9.2009
\begin{center}
{\Large Hadron mass generation and the strong interaction}
\vspace{0.3cm}

H.P. Morsch\\
Institute for Nuclear Studies, Pl-00681 Warsaw, Poland$^*$ 
\end{center}

\begin{abstract}
Based on a Lagrangian with a coupling of two gluons to $J^{\pi}=0^+$
(the quantum numbers of the vacuum) which decay to $q\bar q$
pairs, a model is presented, in which hadrons 
couple directly to the absolute vacuum of fluctuating gluon fields. 
By self-consistency requirements the confinement potential as well as
$q\bar q$ densities and masses are obtained, which are in good 
agreement with experimental data on scalar and vector mesons. 
In comparison with potential models additional states are predicted,
which can explain the large continuum of scalar mesons in the low mass
spectrum and new states detected recently in the charm region.\\
The presented model is consistent with the concept, that the hadron
masses can be understood by binding effects of the quarks.  

PACS/ keywords: 12.38.Aw, 12.38.Lg, 12.39.Mk/ Gluon-gluon coupling to
$q\bar q$-pairs, generation of stable $q\bar q$-systems  
assuming massless quarks, confinement potential, masses of $0^{++}$ and
$1^{--}$ mesons. 
\end{abstract}

In the hierarchy of quantum systems hadrons represent the smallest
complex substructures known inside of atoms and nuclei. This is
supported by the property of asymptotic freedom~\cite{GW} of the
strong interaction. Therefore, hadrons can be related directly to the
absolute vacuum of fluctuating gluon fields (with average energy $\bar
E_{vac}=0$) if the quark masses are zero. 
In quantum chromodynamics (QCD) the vacuum is more complex with a finite
mass of the quarks, which may be coupled to scalar Higgs particles,
subject of extensive searches with the available and planned 
high energy experiments at Fermilab and CERN. 

To investigate hadron mass generation we start from a Lagrangian,
in which two gluons couple to scalar fields, from which $q\bar q$
pairs are emitted. The possibility that two gluons coupled to $0^+$
may produce colour singlet bound states has been mentioned already by
Cornwall~\cite{Corn}, but in context to the structure of QCD. Since
the overlap of gluon fields is non-local, such a model could serve
also as an effective theory to 
investigate the strong scalar fields observed in hadron
excitations~\cite{MoZuneu} and scattering~\cite{Mach}--\cite{pom1},
which may be difficult to extract from QCD. 

Assuming a scalar coupling of gluon fields of the form $g g
\rightarrow (q\bar q)^n$ we write the Lagrangian in the form 
\begin{equation}
\label{eq:Lagra}
{\cal L}_{SI}=\bar \Psi\ i\gamma_{\mu}D^{\mu}\ \Psi -\frac{1}{4} \large(
F_{\mu\nu}F^{\mu\nu} - G_{\mu}G^{\mu}) \ ,
\end{equation}
with $D^{\mu}$ being the covariant derivative
$D^{\mu}=\partial^{\mu}-ig_s A^{\mu}$ and $F^{\mu\nu}$ the 
Abelian field strength tensor
$F^{\mu\nu}=\partial^{\mu}A^{\nu}-\partial^{\nu}A^{\mu}$,  
with $A^{\mu}$ being the gluon fields. $G_{\mu}G^{\mu}$
couples two gluon fields with $J^{\pi}=0^+$ to $q\bar q$-pairs with
\begin{equation}
\label{eq:scalar}
G^{\mu}= - {g_s} A^{\mu}\ [b_1(\bar \Psi V_{1g} \Psi)   
+ b_2 (\bar \Psi V_{1g} \Psi)^2+  b_3 (\bar \Psi V_{1g} \Psi)^3+...] \ ,
\end{equation}
where $(\bar \Psi V_{1g} \Psi)$ stands for the creation of $q\bar q$-pairs,
which interact by gluon exchange. Corresponding Feynman diagrams are 
shown in the upper part of fig.~1. 

The structure of ${\cal L}_s=\frac{1}{4} G_{\mu}G^{\mu}$ 
implies a colour neutral coupling of two gluon fields. Hence, the
symmetry is simply isospin SU(2) (two quarks with different
charge and one gluon)  without colour and (because of massless
quarks as seen below) without flavour degree of freedom. By the coupling of two
gluons to $J^{\pi}=0^+$ the Lagrangian has no chiral symmetry, leading
naturally to the sequence of hadronic states as observed experimentally. 

A two-gluon field is produced only, if there is spacial
overlap of two gluon fields. We write the radial part of ${\cal L}_s$
by a matrix element   
\begin{equation}
\label{eq:phi}
\Phi(r_1-r_2)=\alpha_s <A_{1}(r_1)|\ [b_1(a^{\dagger}_qV_{1g}
a_q) + b_2 (a^{\dagger}_qV_{1g} a_q)^2+ ...]^2\ |A_{2}(r_2)>\ ,  
\end{equation}
where  $A_i(r)$ are radial gluon fields, $a^{\dagger}_q$ and $a_q$
quark and antiquark creation operators, and $V_{1g}$ an interaction
dominated by 1-gluon exchange between these
quarks\footnote{More-gluon exchange as well as spin-spin and
  spin-orbit effects have not been considered.}. For the following
discussion only the first term in eq.~(\ref{eq:phi}) is
needed, contributions due to $(q\bar q)^2$ contributions will be
discussed briefly at the end of the paper. 

Due to the non-local 2-gluon field, given by a 2-gluon density
$\rho_\Phi(r)$, the recoiling local quark fields are also smeared out
and $a^{\dagger}_qV_{1g} a_q$ can be written as a q-q
potential $V_{qq}(r)$ described by folding a $q\bar q$-density with
the gluon-exchange interaction. The fact, that  $V_{qq}(r)$ is a scalar
potential but the created $q\bar q$-pair has negative parity,
requires a p-wave $q\bar q$-density $\rho^p_{q\bar q}(\vec r)=\rho_{q\bar
  q}(r) \ Y_{1,m} (\theta,\phi)$ leading to    
\begin{equation}
\label{eq:vqq}
V_{qq}(r)= \int dr'\rho^p_{q\bar q}(\bar r')\ Y_{1,m}
(\theta',\phi') \ V_{1g}(r-r')\ , 
\end{equation}
where $V_{1g}(r)$ can act only within the density
$\rho_\Phi(r)$. Therefore the interaction has to be cut towards large
radii and we use the following form 
\begin{equation}
\label{eq:veff}
V_{1g}(r)=(-\alpha_s/r)\ e^{-cr}
\end{equation}
with cut-off parameter $c$ determined in a self-consistent fit of the
2-gluon density. Note, that eq.~(\ref{eq:phi}) indicates a departure from a
purely relativistic description with a lifetime of the created system
$\Delta t$. Causality is fulfilled, if $\Delta t > (r-r')/c$. The
shape of the folding potential~(\ref{eq:vqq}) has been calculated by
multiplying the density and potential in momentum space and
retransforming the product to r-space.

The p-wave character of the $q\bar q$-density gives rise to the
constraint $<r_{q\bar q}>\ =\int d\tau\ r \rho_{q\bar q}(r)=0$ and thus  
\begin{equation}
\label{eq:spur}
\rho_{q\bar q}(r)=\sqrt{3}\ (1+\beta\cdot d/dr)\ \rho_{\Phi}(r)\ , 
\end{equation}
where $\beta$ is determined from the condition $<r_{q\bar q}>=0$.
A consequence of eq.~(\ref{eq:phi}) is, that the radial dependence of
$V_{qq}(r)$ should be the same as $\rho_\Phi(r)$ 
\begin{equation}
\label{eq:rhov}
\rho_\Phi(r)=V_{qq}(r)\ .
\end{equation}

From the different relations between densities and potential in
eqs.~(\ref{eq:vqq}), (\ref{eq:spur}) and (\ref{eq:rhov}) the two-gluon 
density is determined. Self-consistent solutions of
eq.~(\ref{eq:rhov}) are obtained assuming a form 
\begin{equation}
\label{eq:wf}
\rho_{\Phi}(r)=\rho_o\ [exp\{-(r/a)^{\kappa}\} ]^2\ \ with\ \ \kappa \sim
1.5\ 
\end{equation} 
and interaction cut-off parameter $c$, which yields a mean square
radius of the effective interaction between 20 and 80 \% larger than
$<r^2_{\Phi}>$. The self-consistency condition is rather strict (see
the lower parts of fig.~1): using a pure exponential form ($\kappa$=1)
a very steep rise of $\rho_\Phi(r)$ is obtained for $r\to 0$ , but an
almost negligible and flat potential, which cannot satisfy
eq.~(\ref{eq:rhov}). Also for a Gaussian form ($\kappa$=2) no
consistency is obtained: normalised to the inner part of
$\rho_\Phi(r)$, the deduced potential falls off more rapidly towards
larger radii than the density. Only by a density with
$\kappa\sim 1.5$ a satisfactory solution is obtained. 
A self-consistent solution is also shown (in r- and momentum
(Q)-space) in fig.~2 with a quantitative agreement for radii  
$>$0.1 fm and momenta $<$2.5 GeV/c . The localisation in space
indicates, that a stationary (mesonic) system is
formed. Interestingly, the emerging quarks can  couple again to two
gluons giving rise to stable glueballs as well.  

In the Fourier transformation to Q-space the process 
$gg\rightarrow q\bar q$ in question is elastic and consequently the
created $q\bar q$-pair has no mass. However, if we take a finite mass
of the created quarks of 1.4 GeV (such a mass has been assumed in
potential models~\cite{qq} for systems 
of similar size), the dashed line in the lower part of fig.~2 is
obtained and no self-consistent solution is possible. Thus, our
solutions require {\bf massless} quarks and consequently the deduced
hadronic systems can be related directly to the absolute vacuum of
fluctuating gluon fields.

\begin{figure}
\centering
\includegraphics [height=18cm,angle=0] {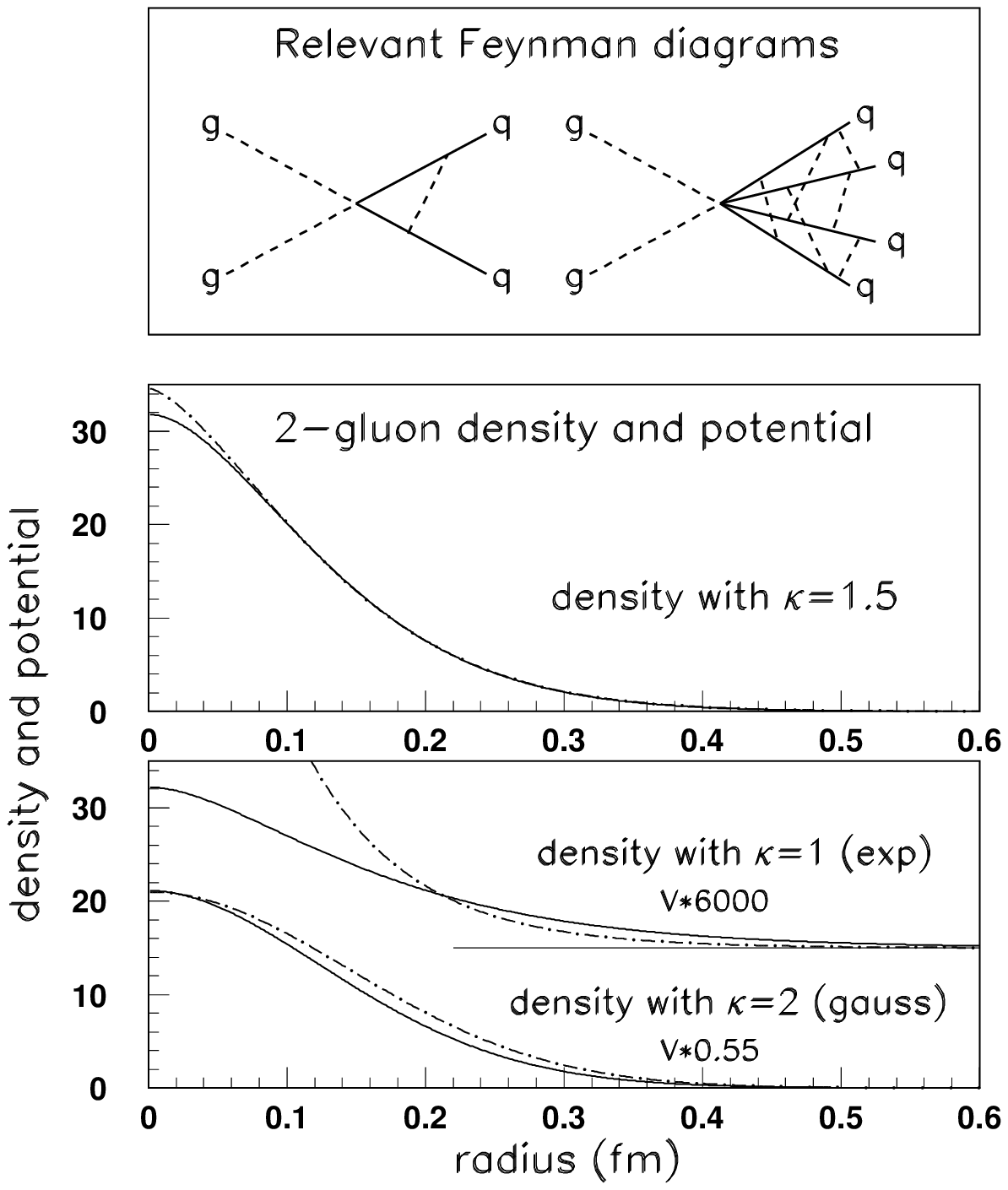}
\caption{Upper part: Relevant Feynman diagrams $2g\to q\bar q$ and
  $2g\to (q\bar q)^2$ (q denotes quark or antiquark).
  Lower parts: density $\rho_\Phi(r)$ with $<r^2>$=0.06 fm$^2$
  (dot-dashed lines) and folding potential~(\ref{eq:vqq}) (solid
  lines) for   $\kappa$=1, 1.5 and 2, respectively.}  
\label{fig:f1}
\end{figure}

\begin{figure} [ht]
\centering
\includegraphics [height=18cm,angle=0] {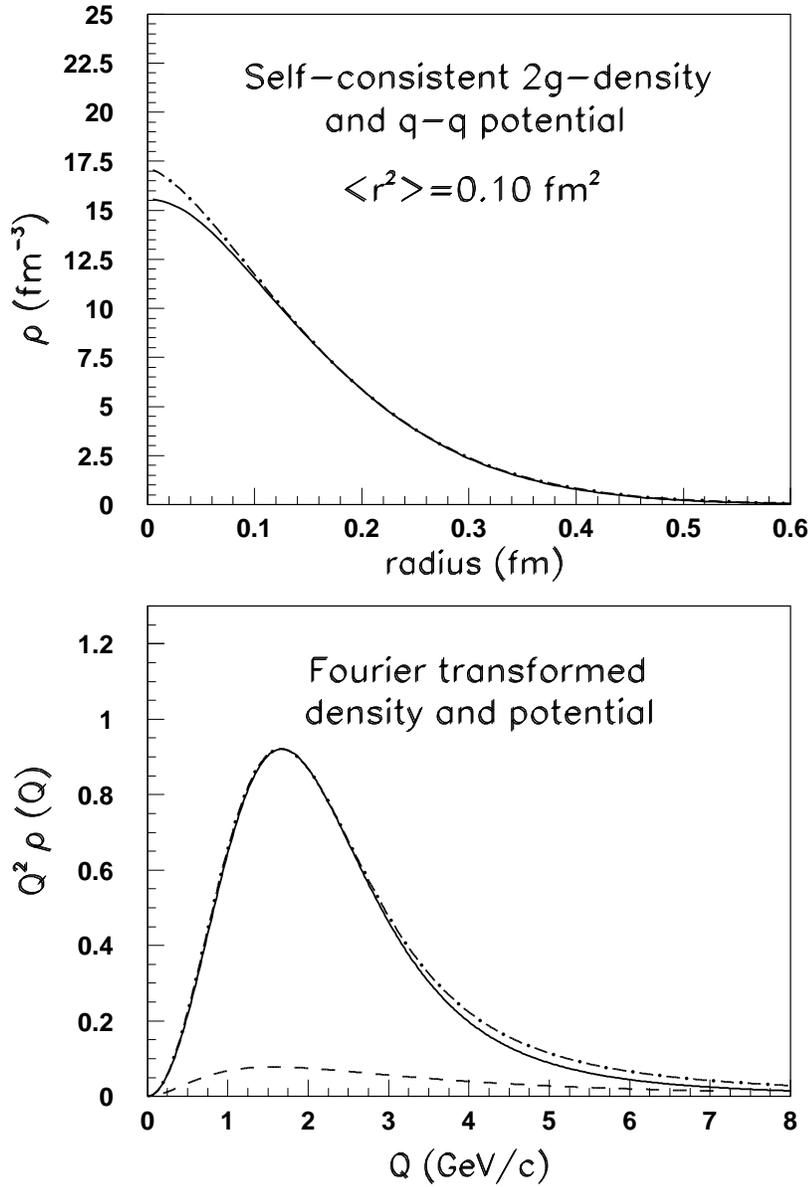}
\caption{Upper part: Self-consistent solution of eq.~(\ref{eq:vqq})
  for a scalar state with two-gluon density and q-q potential given by
  dot-dashed and solid 
  lines, respectively. Lower part: Same as in the upper part but
  transformed to $Q$-space (multiplied by $Q^2$). The dashed line
  corresponds to a calculation assuming quark masses of 1.4 GeV.} 
\label{fig:simdens}
\end{figure}

\begin{figure}
\centering
\includegraphics [height=18cm,angle=0] {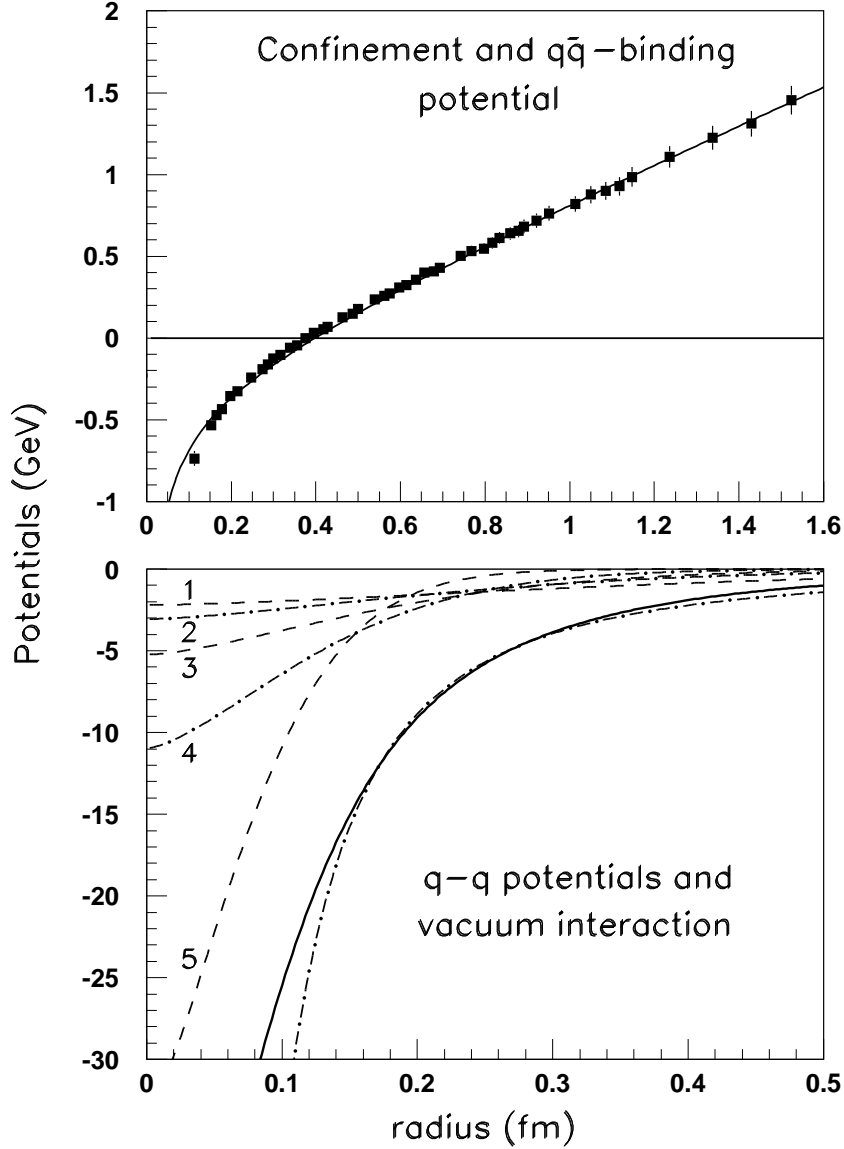}
\caption{Upper part: Confinement potential from lattice
  calculations~\cite{Bali} in comparison with the
  $q\bar q$ binding potential~(\ref{eq:vbind}) given by solid line. 
Lower part: q-q potentials for the different solutions in table~1 with
their sum given by solid line compared to the vacuum sum
rule~(\ref{eq:sum}), lower dot-dashed line.}  
\label{fig:confine}
\end{figure}

For a quark-gluon system with finite mean square radius as shown in 
fig.~\ref{fig:simdens} the corresponding binding potential can be 
obtained from a three-dimensional reduction of the Bethe-Salpeter
equation in form of a (relativistic) Schr\"odinger equation   
\begin{equation}
-\bigg( \frac{\hbar^2}{2\mu_{\Phi}}\ \Big [\frac{d^2}{dr^2} +
  \frac{2}{r}\frac{d}{dr}\Big ] - V_{\Phi}(r)\bigg) \psi_{\Phi}(r) =
  E_i\ \psi_{\Phi}(r)\ ,  
\label{eq:4}
\end{equation}
where $\mu_{\Phi}$ is the mass parameter, which is related to the
mass $m_{\Phi}$ of the $q\bar q$ system by $\mu_{\Phi}=\frac{1}{2}\
m_{\Phi}-\delta m$ with a correction $\delta m\neq 0$
only for very light systems. Further, $\psi_{\Phi}(r)$ is the wave
function, which is given for a system of uncorrelated gluons by
$|\psi_{\Phi}(r)|^2=\rho_{\Phi}(r)$. The dependence of the potential as a
function of $\psi_{\Phi}(r)$ is then given by
\begin{equation}
V_{\Phi}(r)= \frac{\hbar^2}{2\mu_{\Phi}}\ \Big (\frac{d^2\psi_{\Phi}(r)}{dr^2} +
  \frac{2}{r}\frac{d\psi_{\Phi}(r)}{dr}\Big )\frac{1}{\ \psi_{\Phi}(r)}+ V_o\ .  
\label{eq:vb}
\end{equation}
Inserting the form of the density in eq.~(\ref{eq:wf}) yields explicitely    
\begin{equation}
V_{\Phi}(r)= \frac{\hbar^2}{2\mu_{\Phi}}\ \Big[\frac{\kappa}{a^2} (\frac{r}{a})
^{\kappa-2}\ [\kappa (\frac{r}{a})^{\kappa} - (\kappa +1)]\Big]+V_o\ .
\label{eq:vbind}
\end{equation}
Since the $(2g-q\bar q)$ system couples to the vacuum, $V_o$ is
assumed to be zero. With the values of $\kappa$ and $a$ from the self-consistent
solutions~(\ref{eq:wf}) and $\mu_{\Phi}\sim$1.5 GeV (which is
determined below also self-consistently) the deduced densities (shown in 
figs.~\ref{fig:f1} and \ref{fig:simdens}) yield a binding potential
given in the upper part of fig.~\ref{fig:confine}. This has the same
form as the known confinement potential $V_{conf}=-\alpha_s/r+br$
deduced from potential models~\cite{qq} and the lattice data of Bali
et al.~\cite{Bali}. Further, the potential~(\ref{eq:vbind}) is the 
same for other solutions of smaller or larger radii and can therefore
be identified with the 'universal' confinement potential. It is important to note, 
that the self-induced confinement potential~(\ref{eq:vbind})
reproduces the  1/r + linear  form without any assumption on its
distance behavior; this is entirely a consequence of the deduced
radial form~(\ref{eq:wf}) of the two-gluon  density. Interestingly,
because of the rather simple confinement mechanism in our approach, a
direct connection to strings (which have been related to 
the linear part of confinement potential) appears possible.   

To determine bound state energies of basic scalar and vector $q\bar q$
states (with $J^{PC}=0^{++}$ and $1^{--}$) binding in the
self-induced confinement potential (\ref{eq:vbind}) but also in the
q-q potential has to be considered, which (different from the
confinement  potential) depends strongly on 
the radius of the density. For smaller radii a deepening of this 
potential is observed, which leads to more strongly bound states.  

We define the mass of the system by the energy to balance binding
(this is also the way hadron masses are observed). This
yields 
\begin{equation}
\label{eq:mass}
M_{i}=-E_{qq}+E_i \ , 
\end{equation}
where $E_{qq}$ and $E_i$ are the binding energies in $V_{qq}(r)$ and
$V_{\Phi}(r)$, respectively (note that the eigenstates of $V_{qq}(r)$
have negative energy).  
 
We apply another constraint (generation of mass by binding),
requiring that the mass $M_o$ of the lowest bound state is equal to
$m_{\Phi}$ (appearing in the mass parameter
$\mu_{\Phi}=\frac{1}{2}m_{\Phi}-\delta m$ with $\delta m\sim$0). With
this condition we find $m_{\Phi}\sim 1$/$<r^2_{\Phi}>^{1.5}$ for
scalar states with a coupling constant $\alpha_s$ decreasing for
smaller systems. For the five systems in table~1 we find values of
$\alpha_s$ of about 1.3, 0.5, 0.4, 0.32 and 0.23. Together with other
solutions we find $\alpha_s(M_o)\approx 1.5\  \alpha_s^{QCD}(Q)$ for $M_o=Q$ up to large
masses (with values of $\alpha_s^{QCD}(Q)$ from the systematics
in ref.~\cite{PDG}). This yields evidence for asymptotic freedom also
in our model. More details will be discussed elswhere.

So far we have no constraint on the radial extent of
the $(2g-q\bar q)$ system, giving rise to continuous mass
spectra. Discretisation is provided by a vacuum potential sum rule
$\Phi_{vac}(r)=\sum_n \Phi^n(r)$, requiring that the sum of
two-gluon matrix elements~(\ref{eq:phi}) (where $\Phi^n(r)=\alpha_s^n
\rho_{\Phi_n}(r)$) is equal to the total 1-gluon exchange force in
the vacuum. For this we take the simplest form, demanding that the
total cut-off by the finite size of the overlapping gluon fields shows
the same $1/r$ dependence as the 1-gluon exchange force. From this we obtain 
\begin{equation}
\label{eq:sum}
\Phi_{vac}(r)=-\frac{\tilde \alpha_s}{r^2}= -\sum_n \alpha_s^n \rho_{\Phi_n}(r) \ . 
\end{equation}
Five self-consistent solutions for scalar states below 50
GeV have been extracted with mean square radii $<r^2_{\ \Phi}>$ of about
0.49, 0.26, 0.12, 0.06, and 0.02 fm$^2$, respectively, and masses
$M_o$ given in table~1. In the lower part of fig.~3 the different solutions
$\Phi^n(r)$ for n=1-5 are shown together with their sum (solid line). This is
compared to the sum rule~(\ref{eq:sum}) with $\tilde \alpha_s$=0.35 (lower
dot-dashed line). Good agreement is obtained, indicating that the sum
rule is fulfilled. Only for radii $<$0.15 fm a deviation is
observed, which is explained by contributions from solutions with
masses significantly larger than 50 GeV.  
 
\begin{table}
\caption{Deduced masses (in GeV) of scalar and vector $q\bar q$
  states in comparison with known $0^{++}$ and $1^{--}$
  mesons~\cite{PDG}. Only the 1s states in the q-q
  potential are given. The radii of the known states of mass
  $M_o$ are fine-tuned to agree with experimental data.}   
\begin{center}
\begin{tabular}{lc|ccc|ccc}
Solution&(meson)&$M_0$& $M_{1}$ &$M_{2}$&$M_0^{exp}$& $M_{1}^{exp}$ &$M_{2}^{exp}$\\ 
\hline
1~~~$0^{++}$&$\sigma$ &0.60 &1.60$\pm$0.2 & 2.30$\pm$0.3 &0.60$\pm$0.1 \\ 
\hline
2~~~$0^{++}$&$ f_o  $ &1.30 &2.15$\pm$0.2 & 2.80$\pm$0.2 &1.30$\pm$0.2 \\ 
~~~~$1^{--}$&$\omega$ &0.78 &1.66$\pm$0.2&2.28$\pm$0.3 & 0.78& 1.65$\pm$0.02 \\ 
\hline
3~~~$0^{++}$&$ f_o  $ &2.30 &3.00$\pm$0.2 & 3.55$\pm$0.2 &2.30$\pm$0.2 \\ 
~~~~$1^{--}$&$\Phi$   &1.00 &1.69 &2.27&1.02 &1.68$\pm$0.02\\ 
\hline
4~~~$0^{++}$& not seen &4.5$\pm$1.0&M$_o$+0.5 &M$_o$+1.0 & --- \\ 
~~~~$1^{--}$&$J/\Psi$  &3.10&3.68 &4.19& 3.097 & 3.686 & (4.160)  \\ 
\hline
5~~~$0^{++}$& not seen &22.0$\pm$8 &M$_o$+0.4 &M$_o$+0.7 & --- \\ 
~~~~$1^{--}$&$\Upsilon$ &9.46 & 9.99 & 10.36& 9.46 & 10.023 &10.355\\
\end{tabular}
\end{center}
\end{table}
Between the scalar states discussed so far and corresponding vector
states there is a direct relation: the p-wave $q\bar q$ density has to
be replaced by the corresponding s-wave density. This leads to
momentum distributions of the  vector states shifted to smaller
momenta (and corresponding smaller masses) than the scalar states. We
obtain mean square radii  $<r^2_{\ \Phi}>$ of about 0.38, 0.24, 0.10,
and 0.04 fm$^2$ and masses given 
in table~1, which are in good agreement with the masses of the strong
$1^{--}$ mesons (together with their radial excitations) of
the (isoscalar) ``flavour families'' $\omega$, $\Phi$, $J/\Psi$ and
$\Upsilon$. The masses of the known $0^{++}$ states are also in general 
agreement with experiment: the lowest $0^{++}$ state corresponds to
the $\sigma(600)$ meson, which has been clearly identified~\cite{Bugg}
as a broad meson resonance in $J/\Psi$-decay. The next $0^{++}$ states
at 1.3 and 2.3 GeV may correspond to the scalar resonances
$f_o(1300)$, see also ref.~\cite{Aniso}, and $f_o(2300)$, whereas the
higher $0^{++}$ states (not found so far) may not be observable in
e$^+$- e$^-$ collision experiments. 

As compared to potential models with finite quark masses, as e.g.~in
ref.~\cite{qq}, we obtain significantly more states, bound states in the
confinement and in the q-q potential. The solutions in
table~1 correspond only to the 1s levels in the q-q
potential, in addition we have calculated Ns levels
for N=2, 3, and 4. Most of these states, however, have a relatively 
small mass below 3 GeV. As the q-q potential is Coulomb like, it creates
a continuum of Ns levels which ranges down in mass to the threshold
region. This continuum should mix with the states in the confinement
potential, giving rise to large scalar phase shifts at low energies,
which are observed experimentally but so far not well understood in other models. 

Concerning masses above 3 GeV, solution 5 of table~1 yields additional
$0^{++}$ 2s and 3s states in the q-q potential at masses of about 12
and 8.8 GeV, respectively, whereas an extra $1^{--}$ 2s state is
obtained (between the most likely 
$\Psi$(3s) and $\Psi$(4s) states at 4.160 GeV and 4.415 GeV)
at a mass of about 4.2 GeV. This state may 
be identified with the recently discovered X(4260), see 
ref.~\cite{PDG}. Corresponding excited states in the
confinement potential~(\ref{eq:vbind}) should be found 
at masses of 4.9, 5.3 and 5.5 GeV with uncertainties of 0.2-0.3 GeV.

By identifying the deduced $q\bar q$ states with known mesons we can
check the overall consistency of our model by the observed widths of
these states. For the lifetime we assume $<\Delta
t>=<r^2_{\Phi}>^{1/2}/c$ and get for the heavier mesons in 
table~1 values which are small compared to the
lifetime deduced from the relatively small experimental widths.
Differently, for the lightest meson in table~1 we obtain a value of
$<\Delta t>$ of 2.4 10$^{-24}$ s corresponding to a width of about
300 MeV, which is less than the width ($\sim$ 500-600 MeV)
extracted~\cite{Bugg} for $\sigma(600)$. However, as the Coulomb like q-q
potential gives rise to a low energy continuum of Ns states, the width
of the lowest bound state in table~1 has to be much smaller,
indicating that also in this case our approach is valid.   

Finally, possible $(q\bar q)^2$ contributions corresponding to the second
term in eq.~(\ref{eq:phi}) are addressed. In the results shown in 
fig.~2 and the upper part of fig.~4 the folding potential deviates
from $\rho_{\Phi}(r)$ only at small radii. This difference could
be filled by a small $(q\bar q)^2$ component. This is supported
by Monte Carlo simulations~\cite{MZ}, in which good agreement between
$\rho_{\Phi}(r)$ and $V_{qq}(r)$ in eq.~(\ref{eq:rhov}) is obtained in
the entire r and $Q$ region by  the inclusion of a $(q\bar q)^2$
contibution. 

In conclusion, a model has been presented, in which hadron masses are
described as bound states of quarks with a very simple structure of
the vacuum. This leads to a good description of the confinement
potential and hadron masses. Other results, including a discussion of
asymptotic freedom and the stability of baryons will be discussed later. 
\vspace{0.3cm}

We thank P. Decowski, M. Dillig (deceased),
A. Kupsc  and P. Raczka among many other collegues for fruitful
discussions, valuable comments and the help in formal derivations. 
Special thanks to P. Zupranski for writing the Monte
Carlo code, for numerous conversations and clarificatons. \\

$^*$postal address: Institut f\"ur Kernphysik,
Forschungszentrum J\"ulich, D-52425 J\"ulich, Germany, E-mail:
morsch@fz-juelich.de 

\newpage

\end{document}